\documentclass[prc,preprint,showpacs,nofootinbib]{revtex4}
\usepackage{graphicx}

\def\bge{\begin{equation}}
\def\ene{\end{equation}}
\def\bg{\begin{eqnarray}}
\def\en{\end{eqnarray}}

\begin{document}

\title{Neutron Fermi Liquids under the presence of a strong magnetic field with effective nuclear forces}

\author{M. \'{A}ngeles P\'{e}rez-Garc\'{i}a~$^{1}$~\footnote{mperezga@usal.es}, 
J. Navarro~$^{2}$~\footnote{navarro@ific.uv.es}, 
A. Polls~$^{3}$~\footnote{artur@ecm.ub.es}}

\affiliation{$^{1}$Departamento de F\'{i}sica Fundamental and IUFFyM,\\ Universidad de Salamanca, Plaza de la Merced s/n 37008 Salamanca\\ 
$^{2}$IFIC (CSIC Universidad de Valencia), Apdo. 22085 E-46071 Valencia\\
$^{3}$Departament d'Estructura i Constituents de la Materia, Facultat de Fisica,\\ Universitat de Barcelona, 08028 Barcelona}

\date{\today}

\begin{abstract}
Landau's Fermi Liquid parameters are calculated  for non-superfluid pure neutron matter in the presence of a strong magnetic field at zero temperature. The particle-hole interactions in the system, where a net magnetization may be present,  are characterized by these parameters in the framework of a multipolar formalism. We use either zero- or finite-range effective nuclear forces to describe the nuclear interaction. Using the obtained Fermi Liquid parameters, the effect of a strong magnetic field on some bulk magnitudes such as isothermal compressibility and spin susceptibility is also investigated. 
\end{abstract}
\vspace{1pc}

\maketitle

\section{Introduction}
\label{intro}

Objects of astrophysical origin such as soft-$\gamma$ repeaters and anomalous X-ray pulsars are believed to be neutron stars with strong magnetic fields and constitute an example of the so called {\it magnetars} \cite{duncan}. On their surface magnetic field strengths can be of the order $B_{magnetar} \approx 10^{14}-10^{15}$ G. According to the scalar virial theorem \cite{lai} allowed field strengths could be, in principle, as big as $B\approx 10^{18}$ G, however, a full detailed study of the gravitational stability condition of the maximum sustainable magnetic field strength in a star remains to be done.

Description of low temperature homogeneous fermion systems was developed by Landau in 1957 \cite{landau} with the characterization of the low energy properties of the system in terms of a set of parameters, i.e., the Landau parameters, which determine the  effective interaction of two quasiparticles at the Fermi surface. Since then, there have been many attempts to characterize nuclear and neutron matter as a Fermi Liquid. Even if the general  framework of the Fermi Liquid Theory (FLT) is well defined, the microscopic derivation of the Landau parameters in the context of nuclear liquids is not yet fully solved and their values depend on many-body effects and details of the nucleon-nucleon (NN) interaction  that need to be treated carefully. Microscopic calculations of Landau parameters for nuclear and neutron matter for realistic interactions have been subject of several investigations \cite{di87,baldo94,lomb03}. These calculations have also been evaluated in the framework of effective interactions, which allow for a Hartree-Fock description of the system and for a clean definition of the quasiparticle excitations. In this context they have been calculated for symmetric and asymmetric nuclear matter \cite{ventura,npa-nav1}.

 From the astrophysical point of view the importance of FLT is due to the fact that it constitutes an excellent  tool to study the properties of matter in the interior of neutron stars where the low temperature approximation is relevant. In the solid outer crust, the picture of non-superfluid neutrons is  adequate, however, at higher densities in the crust one should consider a finite pairing gap energy \cite{superfluid}. In this non-superfluid scenario one could approximate, at these densities, the fermion system in the very neutron rich astrophysical object as formed by pure homogeneous neutron matter as that described by the FLT. Due to the presence of strong magnetic fields in these objects the possibility of magnetized matter must be taken into account.

In a previous contribution we analyzed the effects of a strong magnetic field of astrophysical origin in a neutron gas~\cite{prc-ang}. There, we found that using different parametrizations for the non-relativistic effective NN interaction  such as Skyrme and Gogny forces, in the context of Hartre-Fock calculations, a net magnetization is energetically allowed. 
One should notice that  effective nuclear interactions such as Skyrme forces predict that even in the absence of a magnetic field an spontaneous magnetization at sufficiently high densities may arise \cite{margueron, rios05,lopez06,isayev}. However, modern calculations using realistic NN potentials such as the Auxiliary Field Diffusion Monte Carlo (AFDMC) method \cite{fantoni}, lowest order constrained variational (LOCV) method  \cite{LOCV}, Brueckner-Hartre-Fock \cite{vida02,vidana}, relativistic mean field \cite{nino} or relativistic  Brueckner-Hartre-Fock \cite{sammarruca}, which have studied  the energetics of spin polarized neutron matter in the absence of magnetic fields, seem to prevent an spontaneous ferromagnetic phase transition and therefore there is the tendency to consider this fact as a pathology of the model used for the effective interaction. Notice also that the modern finite range Gogny effective  interactions do not predict this instability \cite{lopez06}. In any case, the transition into a  ferromagnetic state could have important consequences for the evolution of a protoneutron star, in particular, for the spin correlations in the medium which do strongly affect the  neutrino opacities inside the star \cite{reddy, vau}. For lower densities, as those in the outer crust, microscopic calculations for the non-homogeneous {\it pasta} phases \cite{pasta, watanabe, maruyama, hor} and the spin response in the presence of strong magnetic fields need to be fully calculated. 

Therefore it is interesting to study neutron matter in the presence of a strong magnetic field or with net polarization. As a first step in the description of the system, besides its energetics, which has already been discussed in a previous publication \cite{prc-ang}, we will study the Landau parameters of the polarized neutron matter, characterized for having two Fermi spheres corresponding to the two possible spin orientations of the neutrons. This study requires the use of  an extension of the FLT to a two component system, which is presented in section \ref{formalism}, together with the derivation  of the Landau parameters for   two parametrizations of the nuclear interaction, namely effective Skyrme and Gogny forces. Using a multipolar expansion of the particle-hole matrix elements we will examine the Landau's Fermi Liquid parameters up to dipolar order. Other magnitudes such as isothermal compressibility and spin susceptibility will be also analyzed. Results will be presented in section~\ref{results}, and summary and conclusions will be given in section~\ref{summary}.  

\section{Polarized neutron matter as a Fermi Liquid}
\label{formalism}
In this section we briefly review the basic formalism of the normal FLT \cite{bookFL} for  a homogenous neutron system in the presence of a strong magnetic field. 

We consider a uniform magnetic field in the z-direction, ${\bf B}=B {\bf k}$ which also defines the quantization axis for  the spin. In such a fermionic system particles can have spin projection on the  z-axis, $\sigma$, which can be either $\sigma=+1$ or  $\sigma=-1$ for spins aligned parallel or antiparallel to the magnetic field, respectively. Let us remind here that for a neutron the magnetic moment is antiparallel to the spin. The neutron number density, $\rho$, is the sum of spin up ($+$) and down ($-$) particles , 
\begin{equation}
\rho=\rho_{+}+\rho_{-}.
\label{dens}
\end{equation}
These densities define the Fermi momenta, i.e., the Fermi surface of each fermion component with  spin projection $\sigma$. At zero temperature the number density for each component is given by (we set $\hbar=c=1$),
\begin{equation}
\rho_{\sigma}=\frac{k^3_{F,\sigma} }{6 \pi^2}.
\label{densig}
\end{equation}
In the neutron system the net magnetization density is defined as, 
\begin{equation}
m=\mu_n \Delta \rho,
\end{equation}
where $\mu_n=-1.9130427(5)\mu_N$ is the neutron magnetic moment in units of the nuclear magneton~\cite{pdb} and $\Delta$ is the spin excess or polarization of the system. 
\begin{equation}
\Delta =\frac{\rho_{+}-\rho_{-}}{\rho}.
\label{delta}
\end{equation}

The total magnetization of a given volume is then $M=\int m dV$. The relevant thermodynamical potential to study neutron matter at zero temperature under an external magnetic field, $H$, is the Helmholtz free energy, $F$, defined as~\cite{callen}  
\begin{equation}
F=E-HM,
\label{Fm}
\end{equation}
where $E$ is the energy of the system. Note that previously we have used $B$ to designate the magnetic field strength, however, the total magnetic field is the sum of the external magnetic field and the induced magnetization, ${\bf B}={\bf H}+4 \pi {\bf M}$. In this work we will assume that the ratio $\mid H/B \mid$ will always be close to unity. In what follows we will keep the notation using $B$ to designate the total magnetic field strength. 
 
In the context of FLT it is assumed that ph excitations happen around the Fermi surface, so, this treatment is valid, in principle, for the low temperature regime where $T<<T_{F,\sigma}$ \cite{bookFL}. Then, there is a one-to-one correspondence of states in the interacting system with states in a free Fermi gas. As particles with given momentum ${\bf k}$ and spin projection $\sigma$ are added addiabatically to the system, an eigenstate of the real gas is obtained and, thus, a distribution of quasiparticles, $n_{{\bf k},\sigma}$.

Since the lifetime of the quasiparticle varies inversely to the square of the departure of its energy from the Fermi energy \cite{bookFL} we will consider that the ph perturbation mechanism takes place around the available polarized Fermi seas.  Asuming small deviations of the distribution function of the $\sigma$-polarized quasiparticles in the plasma, $n_{{\bf k},\sigma}$,  with respect to the ground state distribution, $n^0_{{\bf k},\sigma}$ we have $\delta n_{{\bf k},\sigma}=n_{{\bf k},\sigma}-n^0_{{\bf k},\sigma}$.

In the FLT the variation of the free energy, $\delta F$, can be written up to secon order in $\delta n_{{\bf k},\sigma}$ as,  
\begin{equation}
\delta F=\sum _{{\bf k},\sigma}(\epsilon_{{\bf k},\sigma}-\mu_n \sigma B) \delta n_{{\bf k}, \sigma}+\frac{1}{2} \sum_{{\bf k},{\bf k}',\sigma,\sigma'} f_{{\bf k},\sigma,{\bf k}',\sigma'} \delta n_{{\bf k},\sigma} \delta n_{{\bf k}',\sigma'}+O(\delta n^3).
\label{Free_ener}
\end{equation}
As seen on \cite{prc-ang} the change in free energy under the presence of a magnetic field is due to the energy of (anti)alignement of spins in the field and to the change in the quasiparticle distribution function. The single particle energy can be obtained as a functional derivative 
\begin{equation}
\epsilon_{{\bf k},\sigma}-\mu_n \sigma B=\frac{\partial F}{\partial n_{{\bf k},\sigma}},
\label{epsilon}
\end{equation}
as well as the quasiparticle interaction coefficients, 
\begin{equation}
f_{{\bf k},\sigma,{\bf k}',\sigma'}=\frac{\partial^2 F}{\partial n_{{\bf k},\sigma}\partial n_{{\bf k}',\sigma'}}.
\label{fcoef}
\end{equation}

In the general case of polarized neutron matter, the ph matrix element can be written as \cite{annals, npa-nav2}
\begin{equation}
V_{ph}({\bf q_1},{\bf q_2},{\bf q})=\langle {\bf q+q_1} \sigma_1, {\bf q_1} \sigma_3|V|{\bf q+q_2} \sigma_4, {\bf q_2} \sigma_2 \rangle,
\label{vph}
\end{equation}
where we keep the same notation as defined in \cite{annals,npa-nav1,npa-nav2,ventura}  to designate ${\bf k_1=q+q_1}$, ${\bf k_2=q_2}$, ${\bf k_3=q_1}$ and ${\bf k_4=q+q_2}$ as the participating momenta of the ph interaction with initial momenta ${\bf q_1, q_2}$ and ${\bf q}$ is the transferred three-momentum. Assuming that the ph perturbation mechanism takes place around each Fermi surface, each quasiparticle incoming momentum, ${\bf q_i}$, must be replaced by the corresponding polarized Fermi momentum ${\bf k_{F,\sigma}}$ and the momentum transfer is supposed to be small ${\bf q} \approx 0$.

When computing the quasiparticle interaction matrix elements we must note that they only depend on the Fermi momentum of each polarized component and the relative angle, $\theta$, of interacting quasiparticles three-momenta ${\bf q_1}$, ${\bf q_2}$. 
For a non polarized system it is usually written  as an expansion in Legendre polinomials~\cite{bookFL} 
\begin{equation}
V_{ph}=\sum_{l=0}^{\infty} \big [ f_l + g_l {\bf \sigma_1 .\sigma_2} \big ] P_l ( cos\theta) ,
\label{elemento}
\end{equation}

Previous works \cite{iwamoto, vau, npa-nav1, npa-nav2} have studied a non polarized pure neutron system using a ph  interaction at the monopolar and dipolar ($l=0, 1$) level. In that case, it is convenient to define dimensionless parameters $F_l=f_l N_0$ and $G_l=g_l N_0$ where $N_0$ is the quasiparticle level density at the Fermi surface at $T=0$. For non polarized neutron matter of density $\rho=g k^3_F/(6 \pi^2)$ is given by $N_0=\frac{g m^{*}k_F}{2 \pi^2}$ where $g=2$ is the spin degeneracy and $m^*$ is the effective mass at the Fermi surface.

Considering the two possible spin orientations $\sigma=\pm 1$ the polarized interaction matrix elements can be written using coefficients depending on the polarizations involved, $f_l^{(\sigma,\sigma')}$. On a non polarized system they fullfill the following relation,
\begin{equation}
f_l=\frac{f_l^{(\sigma,\sigma)}+f_l^{(\sigma,-\sigma)}}{2},
\label{f0sum}
\end{equation}
\begin{equation}
g_l=\frac{f_l^{(\sigma,\sigma)}-f_l^{(\sigma,-\sigma)}}{2}.
\label{g0sum}
\end{equation}
In the same way, as mentioned for the non polarized case, one can define dimensionless coefficients in the form $F^{(\sigma, \sigma')}_l=f^{(\sigma, \sigma')}_l \sqrt{N_{0 \sigma} N_{0 \sigma'}}$. where $N_{0 \sigma}$ is the  quasiparticle level density at the Fermi surface of each component at $T=0$, $N_{0 \sigma}=\frac{m^{*}_{\sigma} k_{F,\sigma}}{2 \pi^2}$.
The Landau coefficients for the non polarized case can be recovered from the general polarized coefficients as $\Delta \rightarrow 0$,
\begin{equation}
F_l^{(\sigma)} \equiv \frac{ F_l^{(\sigma,\sigma)}+ F_l^{(\sigma,-\sigma)}}{2} \rightarrow \frac {F_{l}}{2} \: (B=0),
\label{flstable1}
\end{equation}
\begin{equation}
G_l^{(\sigma)} \equiv \frac{ F_l^{(\sigma,\sigma)}- F_l^{(\sigma,-\sigma)}}{2}\rightarrow \frac {G_{l}}{2} \: (B=0),
\label{glstable1}
\end{equation}
where we have defined $F_l^{(\sigma)}$ and $G_l^{(\sigma)}$. Note that some works define combinations of coefficients in a different way \cite{n0}. Contrary to the non polarized case, now the $F_l^{(\sigma,\sigma')}$ interaction displays a $2 \times 2$ matrix structure depending on the spin projections considered. To size the importance of the polarization in neutron matter it is useful to consider a ratio of  these coefficients $(F_l^{(\sigma)}, G_l^{(\sigma)})$ with respect to the $B=0$ case. We define 
\begin{equation}
R_{F l}^{\sigma}=\frac{2 F_l^{(\sigma)} - F_{l}(B=0)}{|F_{l}(B=0)|},
\label{rf}
\end{equation}
and
\begin{equation}
R_{G l}^{\sigma}=\frac{2 G_l^{(\sigma)} - G_{l}(B=0)}{|G_{l} (B=0)|}.
\label{rg}
\end{equation}
In this work we will concentrate on $R_{F l}^-$, $R_{G l}^-$ which correspond to the energetically favourable dominant population fraction with magnetic moments (spins) aligned parallel (antiparallel) to the magnetic field.

In this work we will be interested in retaining just the terms with $l \leq 1$. Notice that constraining results just to dipolar multipolarity is exact for the zero-range Skyrme forces but for the finite-range Gogny forces the coefficients at higher multipolarities are not zero although they decrease rather rapidly. 

Within the context of the Landau FLT, additional static physical quantities of interest \cite{ostgaard,2dimgas} can be obtained. We have derived the expressions for these quantities for the polarized neutron matter case. The quasiparticle effective mass is related via Galilean invariance with the dipolar matrix elements \cite{bookFL},
\begin{equation}
m^*_{\sigma}/m=1+\frac{1}{3} N_{0 \sigma} \big [ f_1^{(\sigma,\sigma)}+(\frac{k^2_{F,-\sigma}}{k^2_{F,\sigma}})f_1^{(\sigma,-\sigma)} \big ].
\label{mef}
\end{equation}
The isothermal compressibility, $K=9 \frac{\partial P}{\partial \rho}$, is related to the study of the variation of the pressure and density profiles for a neutron star. At zero temperature can be written as,
\begin{equation}
K=\frac{9}{\rho} \sum_{\sigma} \frac{\rho^2_{\sigma}} {N_{0 \sigma}}  \big ( 1+N_{0 \sigma} \big [ f_0^{(\sigma,\sigma)}+(\frac{k^2_{F,-\sigma}}{k^2_{F,\sigma}})f_0^{(\sigma,-\sigma)} \big ] \big ).
\label{kappa}
\end{equation}
The thermodynamical magnitude that characterizes the change in the magnetization of the medium when an external magnetic field exists is the spin susceptibility, $\chi$. It is obtained by considering the change of the quasiparticle distribution functions for the up and down components \cite{vidaurre},
\begin{equation}
m=\chi B=\mu_n (\delta \rho_+ -\delta \rho_-).
\label{magnet}
\end{equation}
For a polarized system we have obtained
\begin{equation}
\chi=\sum_{\sigma} \frac{\mu_n^2 N_{0 \sigma}} { 1+N_{0 \sigma} \big [ f_0^{(\sigma,\sigma)}-(\frac{k^2_{F,-\sigma}}{k^2_{F,\sigma}})f_0^{(\sigma,-\sigma)} \big ] \big )}.
\label{chi}
\end{equation}
The bulk magnitudes, $K$ and $\chi$ for polarized neutron matter are affected by a factor involving the quasiparticle interaction matrix elements and the level densities with respect to the polarized free Fermi gas value for each population component.

A necessary ingredient in the calculation of the free energy functional in the neutron system, $F$, is the nuclear interaction. In the next subsections of the present paper we consider either zero-range Skyrme or finite-range Gogny forces as illustrative examples of effective nuclear interactions. Using the non-relativistic Hartree-Fock approximation, the free energy per particle in neutron matter under the presence of a magnetic field $B$ can be calculated minimizing $F$ as a function of $B$ for given thermodynamical conditions to  obtain $\Delta_{min}$  (polarization at the minimum). As a rule of thumb, for the maximum magnetic field strength considered in this work, $B=10^{18}$ G, a $\Delta_{min}<40\%$ is allowed for densities $0.5<\rho/\rho_0<3.3$ at zero temperature \cite{prc-ang}. 

\subsection{Skyrme force}
\label{Skyforce}

We have considered, in first place, the phenomenological Skyrme interaction that appears in the literature under the rather general form \cite{skyrme}
\begin{eqnarray}
V^{Skyrme}_{NN}({\bf r}_1,{\bf r}_2)&=& t_0 \left(1+x_0 P^{\sigma} \right) \delta({\bf r})  + \frac{1}{2} t_1 \left(1+x_1 P^{\sigma} \right)
\left[ {\bf k'}^2 \delta({\bf r}) +  \delta({\bf r}) {\bf k}^2 \right] 
\nonumber \\
&& +  t_2 \left(1+x_2 P^{\sigma} \right) {\bf k'} \cdot \delta({\bf r}) {\bf k}
+ \frac{1}{6} t_3 \left(1+x_3 P^{\sigma} \right) \rho^{\alpha} ({\bf R})\delta({\bf r}) ,
\label{skyrme}
\end{eqnarray}
where ${\bf r}={\bf r}_1-{\bf r}_2$, ${\bf R}=({\bf r}_1+{\bf r}_2)/2$ and 
${\bf k}=({\bf \nabla}_1-{\bf \nabla}_2)/2i$ is the relative momentum acting on the right and ${\bf k'}$ its conjugate acting on the left. $P^{\sigma}$ is the spin exchange operator. Note that we have omitted the spin-orbit term not relevant for the total energy of homogeneous systems. Its effect on the RPA response function has been shown to be sizeable only at values of the momentum transfer much higher than the Fermi momentum \cite{rpa}.

As widely known, this effective interaction allows for a good reproduction of finite nuclei observables and bulk matter equation of state (EOS) relevant to neutron stars~\cite{skyrme}. In the Skyrme model the ph interaction matrix elements retain the density and polarization dependence and their multipolar decomposition gives contribution up to $l=1$. We have derived analytical expressions for both parallel and antiparallel coefficients for an arbitrary polarization in neutron matter. 

The monopolar terms  can be written as,
\begin{eqnarray}
f_0^{(\sigma,\sigma)} &=& \frac{1}{6} t_3 (1-x_3) [\alpha (\alpha-1) \rho^{\alpha-2} \rho_+ \rho_+ + 2 \alpha \rho^{\alpha-1}( \rho - \rho_{\sigma} )]- 
\nonumber \\
&& + t_2 (1+x_2) k^2_{F,\sigma},
\end{eqnarray}
\begin{eqnarray}
f_0^{(\sigma,-\sigma)} &=&
t_0 (1-x_0) + \frac{1}{6} t_3 (1-x_3)\big [ \alpha (\alpha-1) \rho^{\alpha-2} \rho_+ \rho_- + (\alpha+1) \rho^{\alpha}\big ] - 
\nonumber \\
&& + \frac{1}{4} [t_1 (1-x_1) + t_2 (1+x_2) ] \left( k^2_{F,\sigma} + k^2_{F,-\sigma} \right),
\end{eqnarray}
and the dipolar terms can be written as,
\begin{equation}
f_1^{(\sigma,\sigma)} = -t_2 (1+x_2) k^2_{F,\sigma},
\end{equation}
\begin{equation}
f_1^{(\sigma,-\sigma)} = -\frac{1}{2} [t_1 (1-x_1) + t_2 (1+x_2) ] k_{F,\sigma}\rm k_{F,-\sigma}.
\end{equation}

In this paper we use the SLy7 parametrization~\cite{skyrme} of the Skyrme force which not only provides good values for binding of nuclei but also a neutron matter EOS in agreement with microscopic calculations obtained using realistic interaction and giving values of maximum neutron star masses around $1.5 M_{\odot}$. In Table~\ref{tab:param} we summarize the values of some observables for symmetric nuclear matter (SNM): saturation density, $\rho_0$, binding energy for symmetric nuclear matter, $a_v$,  symmetry energy, $a_s$,  and incompressibility modulus, $K_{\infty}$, for the effective interaction models used in this work.
\begin{table}[htbp]
\begin{center}
\caption{Values of some observables for symmetric nuclear matter  in absence of magnetic field with the Skyrme and Gogny forces considered in this work~\cite{skyrme,d1p}.}
\label{tab:param}
\begin{tabular}{l|llll} \hline
Model &$\rho_0 (fm^{-3})$   &$K_{\infty}$ (MeV) & $a_v$ (MeV) & $a_s$ (MeV)\\ \hline
SLy7 &0.158  &229.7 &-15.89  &31.99 \\
D1P &0.1737  &266  &-16.19 &34.09
\end{tabular}
\end{center}
\end{table}

\subsection{Gogny force}
\label{gogny}

The Gogny force has also been extensively used in the literature. In this paper we use the D1P parametrization \cite{d1p} which allows for  a good description of both finite nuclei and EOS of pure neutron matter \cite{lopez06}. It includes a sum of two Gaussian shaped terms that mimic the finite range effects of a realistic interaction in the medium. Usually, it also contains a density-dependent zero range term. The interaction potential is written as:
\begin{eqnarray}
V^{Gogny}_{NN}({\bf r}_1,{\bf r}_2)&=& \sum_{i=1}^{2} \{ [W_i + B_i P^{\sigma} -
H_i P^{\tau} - M_i P^{\sigma} P^{\tau} ] e^{ -|{\bf r}_1-{\bf r}_2|^2 / \mu_i^2 }+ \nonumber \\ 
&& t_{3i}(1+x_{3i} P^{\sigma} ) \rho^{\alpha_i}\delta({\bf r})  \},
\label{v_gogny}
\end{eqnarray}
where $P^{\sigma}$ and $P^{\tau}$ are the spin and isospin exchange operators. The values of some observables for SNM with the Gogny D1P force appear in Table I. We have derived the expressions of the monopolar and dipolar Landau parameters with the Gogny interaction: 
\begin{eqnarray}
f_0^{(\sigma,\sigma)} &=& \sum_{i=1,2} (W_i-H_i+B_i-M_i) \pi^{3/2} \mu_i^3 [ 1-\frac{1}{k^2_{F,\sigma} \mu_i^2} (1-e^{-k^2_{F,\sigma} \mu_i^2}) ]+ 
\nonumber \\
&& t_{3i} (1-x_{3i}) [2 \alpha_i \rho^{\alpha_i-1}(\rho-\rho_\sigma) + \alpha_i (\alpha_i-1)\rho^{\alpha_i-2}  \rho_+ \rho_-],
\label{f0g1}
\end{eqnarray}
\begin{eqnarray}
f_0^{(\sigma,-\sigma)} &=& \sum_{i=1,2}(W_i-H_i) \pi^{3/2}\mu_i^3-
\nonumber \\
&&(B_i-M_i) \frac{\pi^{3/2}\mu_i} {k_F(\sigma) k_F(-\sigma)} [ e^{-\frac{1}{4}(k_{F,\sigma}-k_{F,-\sigma})^2 \mu_i^2} -e^{-\frac{1}{4}(k_{F,\sigma}+k_{F,-\sigma})^2 \mu_i^2} ]+   
\nonumber \\
&& t_{3i} (1-x_{3i})[(\alpha_i +1) \rho^{\alpha_i} + \alpha_i (\alpha_i-1)\rho^{\alpha_i-2}  \rho_+ \rho_- ],
\label{f0g2}
\end{eqnarray}
and for the $f_1$ terms,
\begin{eqnarray}
f_1^{(\sigma,\sigma)} &=& \sum_{i=1,2} -(W_i-H_i+B_i-M_i) \frac{3 \pi^{3/2} \mu_i}{k^2_{F,\sigma}} [1-\frac{2}{k^2_{F,\sigma} \mu_i^2}+ \nonumber \\
&&(1+\frac{2}{k^2_{F,\sigma} \mu_i^2} ) e^{- k^2_{F,\sigma} \mu_i^2}) ],
\label{f1g1}
\end{eqnarray}
\begin{eqnarray}
f_1^{(\sigma,-\sigma)} &=& \sum_{i=1,2} -(B_i-M_i) \frac{3 \pi^{3/2}\mu_i} {k_{F,\sigma} k_{F,-\sigma} } 
[(1-\frac{2} {k_{F,\sigma} k_{F,-\sigma} \mu_i^2}) e^{-\frac{1}{4}(k_{F,\sigma}-k_{F,-\sigma} )^2 \mu_i^2} +\nonumber\\
&&(1+\frac{2} {k_{F,\sigma} k_{F,-\sigma}\mu_i^2}) e^{-\frac{1}{4}(k_{F,\sigma}+k_{F,-\sigma})^2 \mu_i^2} ].
\label{f1g2}
\end{eqnarray}
It is worth mentioning at this point that, once the ph interaction matrix elements have been calculated, the response functions of neutron matter can be obtained along the lines described in \cite{rpa}. 

\section{Results}
\label{results}

In this section we discuss the properties of polarized neutron matter at zero temperature under the presence of a strong magnetic field. Such properties are described in terms of ($l=0,1$) Landau parameters, which are calculated from Skyrme and Gogny effective interactions. At each thermodynamic condition $(\rho, T=0, B)$ the corresponding induced magnetization of the neutron system has been obtained by minimizing the free energy within a Hartree-Fock calculation, as described in \cite{prc-ang}. In order to be physically meaningful we have considered that the nucleon picture holds up to a maximum limiting density $\rho=4\rho_0$ $(\rho_0=0.1737 fm^{-1})$. Similarly, more detailed calculations are needed when considering densities below $\rho=0.5\rho_0$ where  nuclear pasta \cite{pasta, watanabe, maruyama, hor} may be present. 

To measure the effects of the magnetic field on the Landau parameters we calculate the variation ratios, $R^{\sigma}_{F l}$ and $R^{\sigma}_{G l}$ as defined in Eqs.(\ref{rf}) and (\ref{rg}) for the dominant neutron component with magnetic moment aligned parallel to the magnetic field. This corresponds to the spin antialigned component due to the fact that $\mu_n<0$.

\begin{figure}
\begin{minipage}[b]{0.5\linewidth} 
\centering
\includegraphics[angle=-90,scale=0.75]{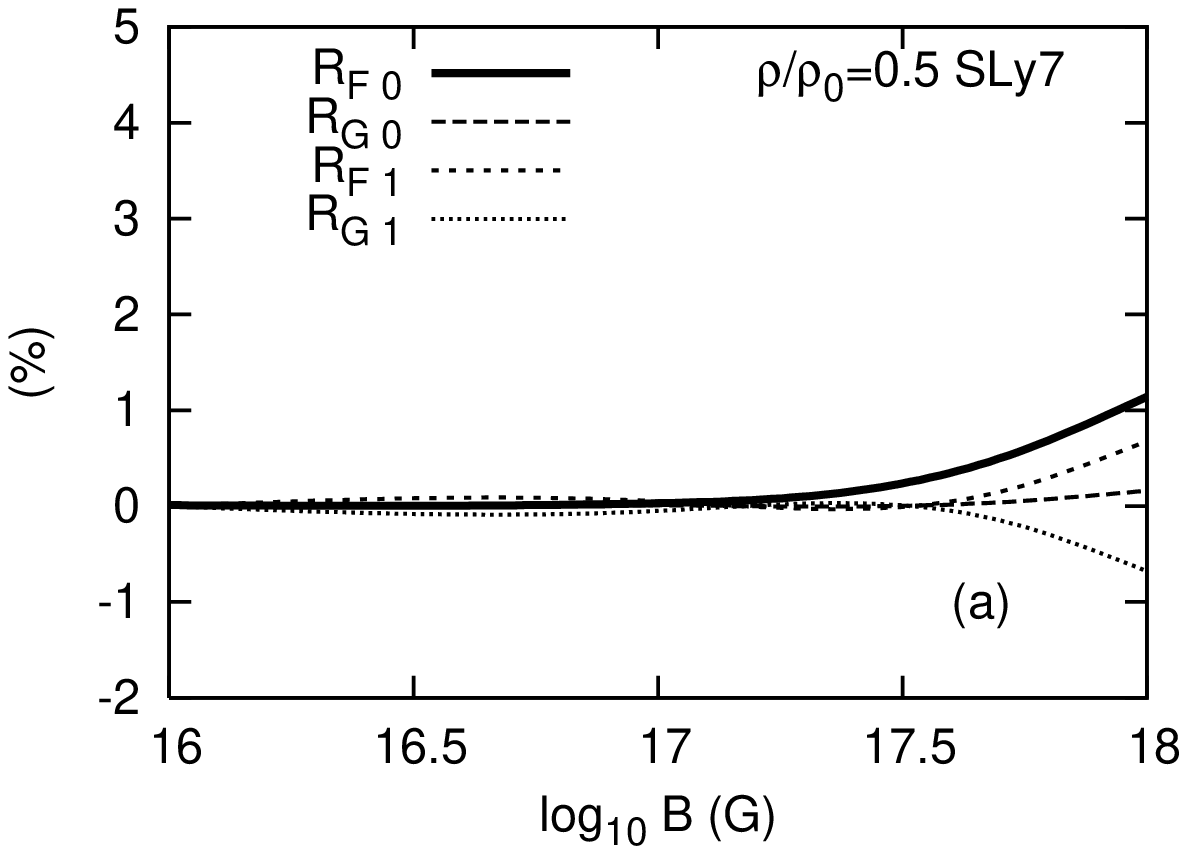}
\end{minipage}
\hspace{0.5cm} 
\begin{minipage}[b]{0.5\linewidth}
\centering
\includegraphics[angle=-90,scale=0.75]{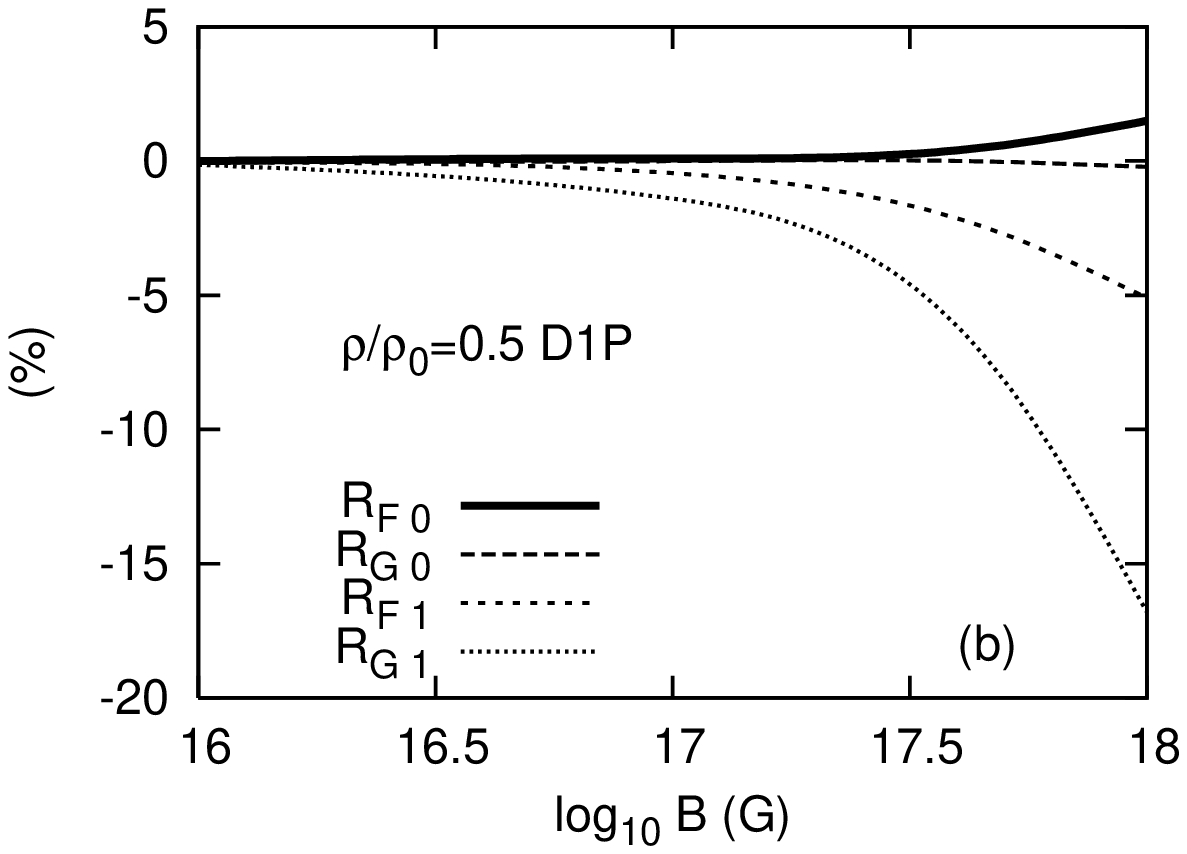}
\caption{Variation ratio of Landau coefficients with the SLy7 (a) and D1P (b) parametrizations at density $\rho=0.5\rho_0$  and  zero temperature as a function of the logarithm of the magnetic field strength.}
\label{Fig1}
\end{minipage}
\end{figure}

In Fig.~\ref{Fig1} we show the ratios $R^{-}_{F 0}$, $R^{-}_{G 0}$, $R^{-}_{F 1}$ and $R^{-}_{G 1}$ (we omit the superscript) computed with the Skyrme SLy7 (a) and with the Gogny D1P (b) interactions for a characteristic low density of $\rho=0.5\rho_0$ as a function of the logarithm of the magnetic field. At this low density the change due to the magnetic field strength in the monopolar and dipolar coefficients is very mild. For Skyrme the ph interactions are mostly slightly more repulsive in the density and spin channel but for Gogny this behaviour is reversed. The variation of the dipolar terms, as computed with the Gogny interaction is larger than for the monopolar case. However, dipolar terms are about one order of magnitude smaller \cite{npa-nav2} so the overall effect is mostly unchanged with respect to the monopolar  description. For magnetic field strengths below $B\approx 10^{16}$ G the change in the ratios is negligible due to the tiny value of the neutron magnetic moment, which results in a vanishing value of the induced magnetization, i. e., of the polarization of the neutron matter.

\begin{figure}
\begin{minipage}[b]{0.5\linewidth} 
\centering
\includegraphics[angle=-90,scale=0.75]{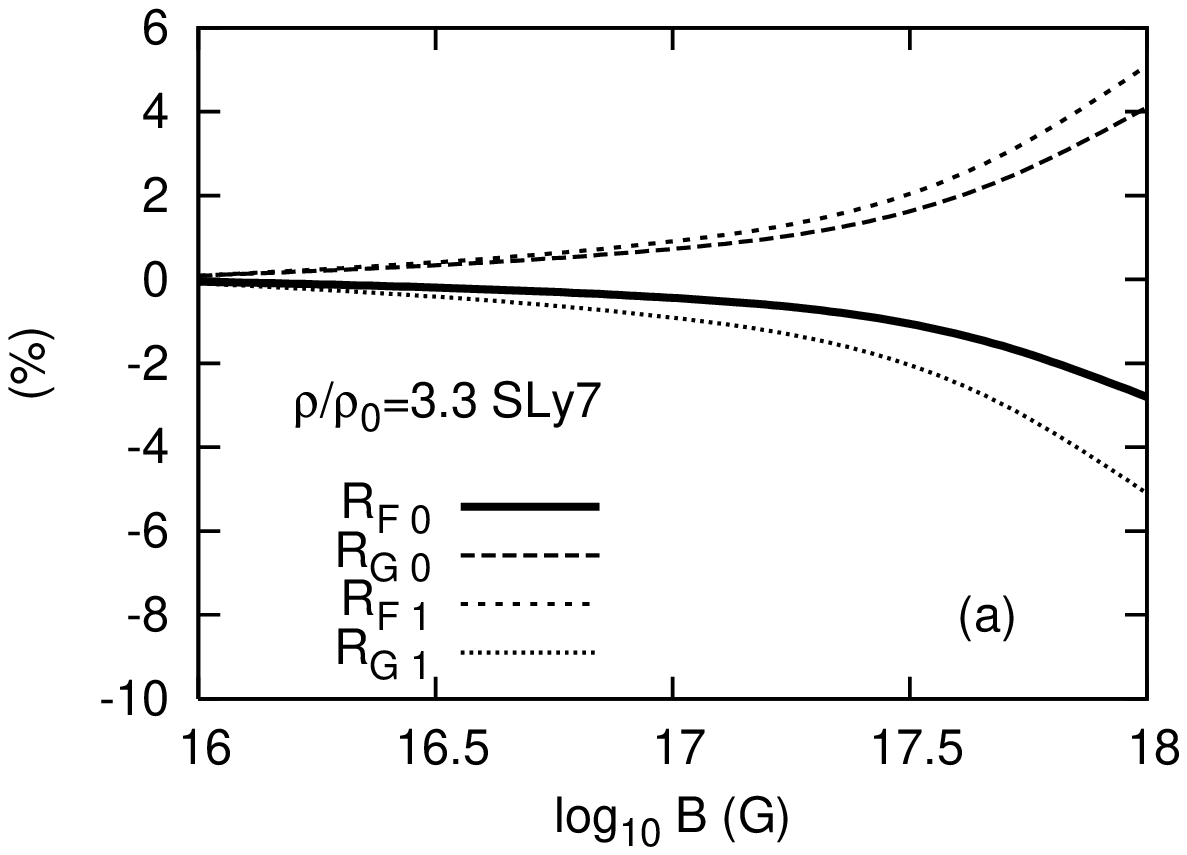}
\end{minipage}
\hspace{0.5cm} 
\begin{minipage}[b]{0.5\linewidth}
\centering
\includegraphics[angle=-90,scale=0.75]{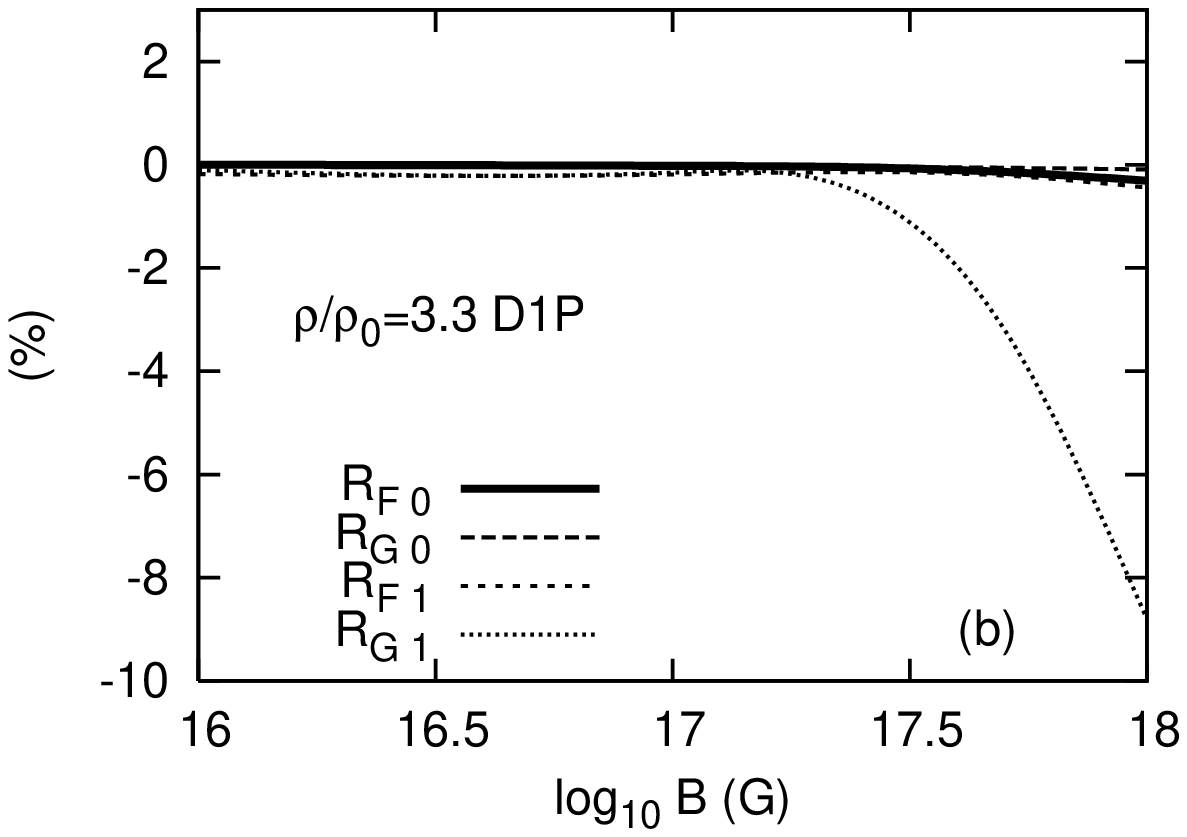}
\caption{Same as Fig.~\ref{Fig1} computed at density $\rho=3.3 \rho_0$.} 
\label{Fig2}
\end{minipage}
\end{figure}

In Fig.~\ref{Fig2} we show the same ratios shown in Fig.~\ref{Fig1} as computed with the Skyrme (a) and Gogny D1P (b) interactions  for a high density case, $\rho=3.3\rho_0$ as a function of the logarithm of the magnetic field. We can see that for the Skyrme SLy7 interaction the presence of a ferromagnetic transition at density close to the one selected \cite{prc-ang}, induces a larger change than at lower densities in the monopolar and dipolar terms as magnetic field grows. The attraction in the ph density excitation channel increases while in the spin channel becomes more repulsive as the magnetic field strength grows. For the dipolar terms this tendency is reversed. For the Gogny interaction the variation of the ratios is very small except for the $G_1$ term. However, the large variation of $G_1$ does not modify the tendency shown by the monopolar terms since in this range of densities it is also an order of magnitude smaller than $G_0$.

\begin{figure}
\begin{minipage}[b]{0.5\linewidth} 
\centering
\includegraphics[angle=-90,scale=0.75]{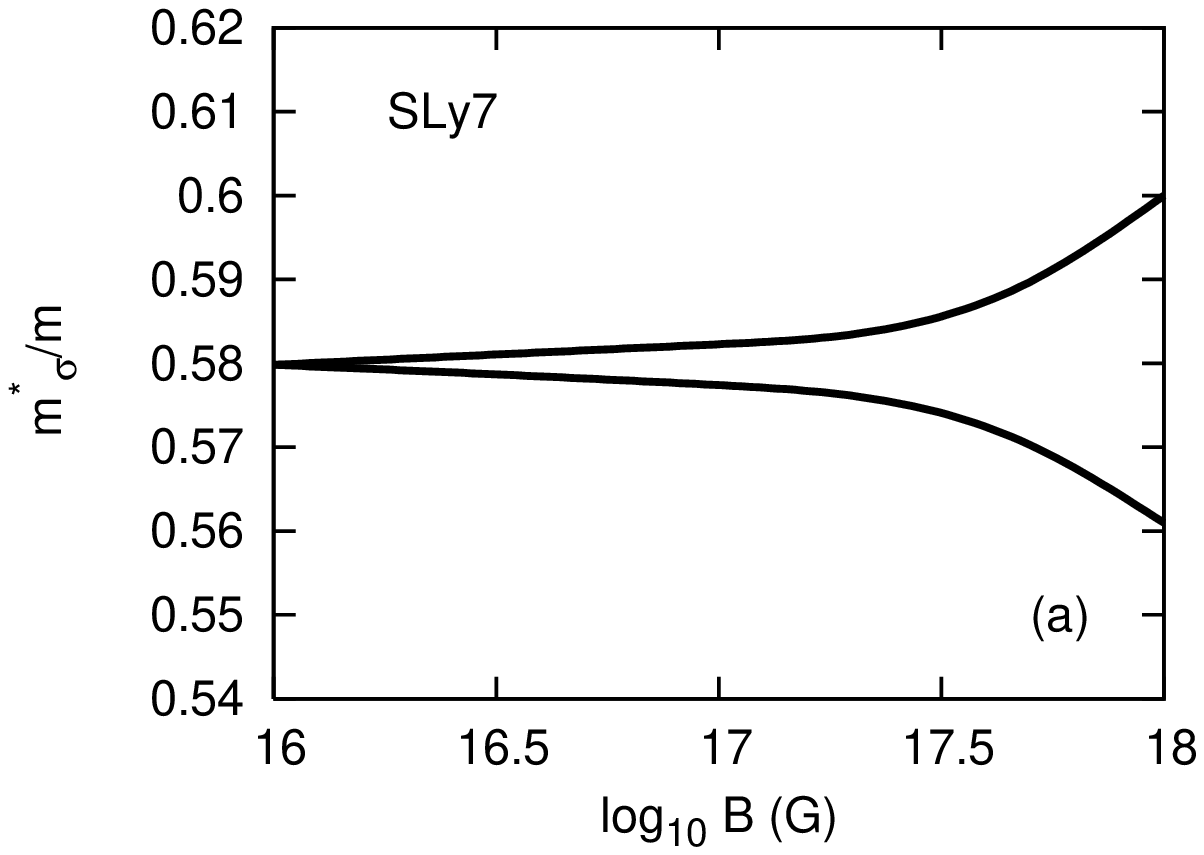}
\end{minipage}
\hspace{0.5cm} 
\begin{minipage}[b]{0.5\linewidth}
\centering
\includegraphics[angle=-90,scale=0.75]{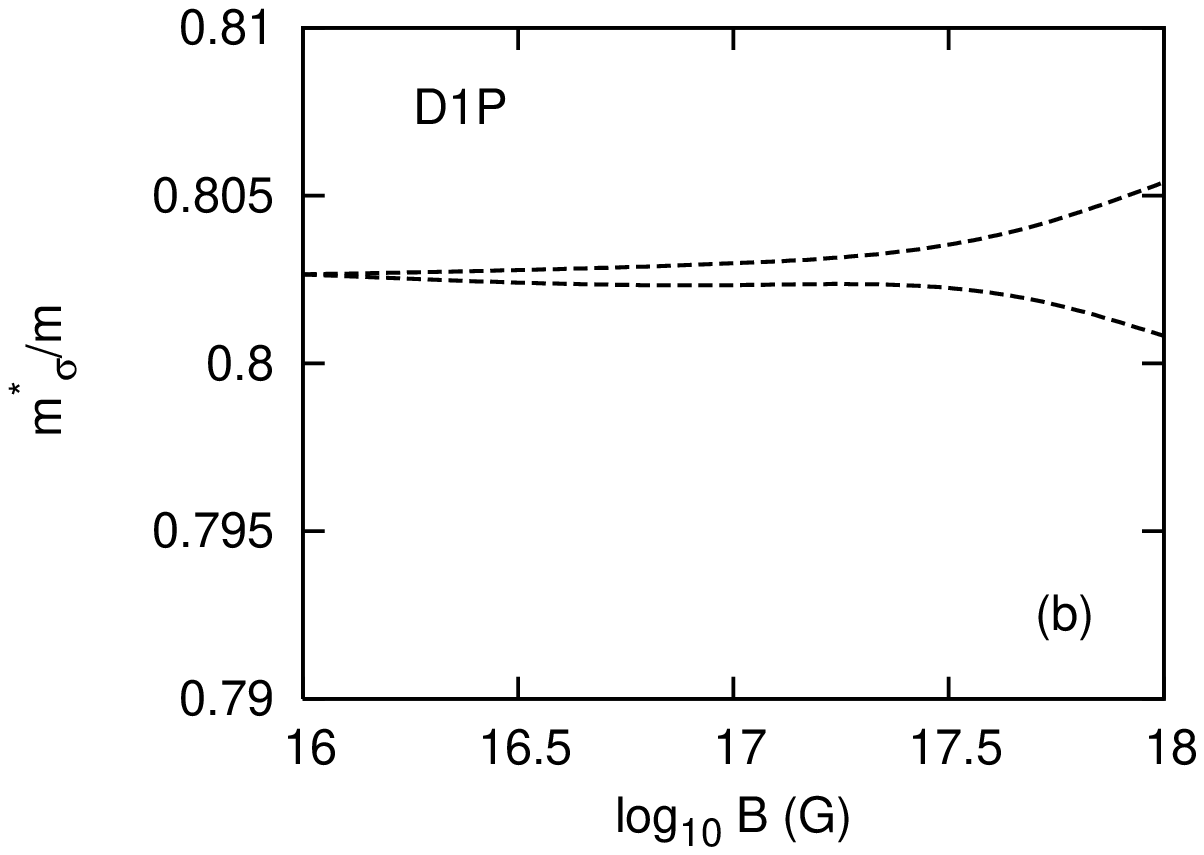}
\caption{Effective neutron mass at density $\rho_0$ as a function of the logarithm of the magnetic field strength for the Skyrme SLy7 (a) and Gogny D1P (b) interactions. For each model upper (lower) curves correspond to spin down (up) polarized particles.} 
\label{Fig3}
\end{minipage}
\end{figure}

We now analyze some other quantities related to the Landau parameters in polarized neutron matter. In Fig.~\ref{Fig3} we show the effective neutron mass for the Skyrme SLy7 (a) and Gogny D1P (b) interactions as a function of the logarithm of the magnetic field strength. In each plot the spin down (upper curve) and spin up (lower curve)  polarized components at saturation density, $\rho_0$, are shown. Skyrme interaction predicts smaller effective masses and a larger relative variation for the spin up and down components than in the Gogny case. These two effects will, in turn, largely affect the level density at the Fermi surfaces of the spin components in magnetized neutron matter.

\begin{figure}[hbtp]
\begin{center}
\includegraphics [angle=-90,scale=.75] {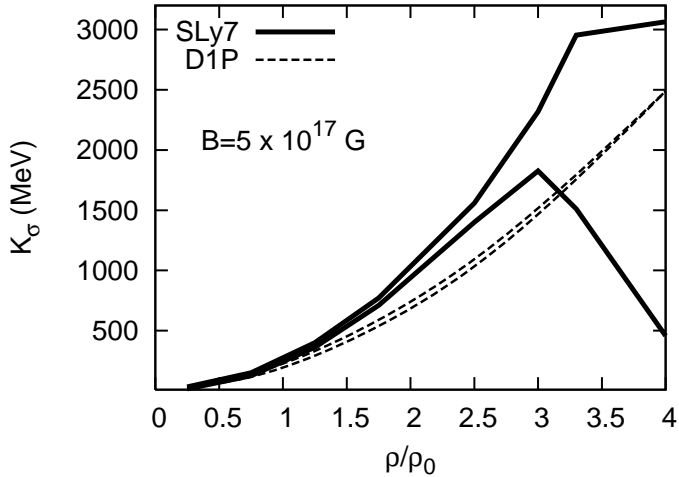}
\caption{Isothermal compressibility components obtained for the Skyrme SLy7 (solid curve) and Gogny D1P (dashed line) models as a function of density. Spin down (upper curve) and up (lower curve) polarized components are shown for each model.}
\label{Fig4}
\end{center}
\end{figure}

\begin{figure}[hbtp]
\begin{center}
\includegraphics [angle=-90,scale=.75] {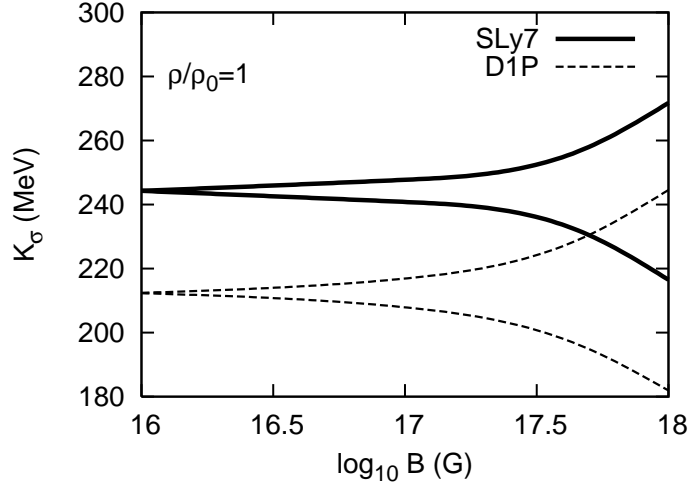}
\caption{Isothermal compressibility components obtained for the Skyrme SLy7 (solid curve) and Gogny D1P (dashed line) models as a function of the logarithm of the magnetic field strength at density $\rho_0$. Upper (lower) curves refer to spin down (up) polarized contributions.} 
\label{Fig5}
\end{center}
\end{figure}

In Fig.~\ref{Fig4} the two spin contributions to the isothermal compressibility for neutron matter are shown as a function of density for a magnetic field strength $B=5 \times 10^{17}$ G and zero temperature for the Skyrme SLy7 (solid line) and Gogny D1P (dashed line) interactions. Contributions from the spin down (up) polarized particles correspond to the upper (lower) curves for each interaction model. In some works compressibility values are plot related to the Fermi gas value, in our case, due to the fact that at each density the relative populations of the up and down spin polarized particles change, we have chosen to plot absolute values. Due to the mild variation of the Landau coefficients with the magnetic field (lower than $20\%$) we can see that the main effect of the magnetization in the system is the change in the level densities. As the density increases the dominant spin down fraction of the polarized plasma becomes stiffer. However, there is a dramatic decrease in the global compressibility in the proximity of the ferromagnetic transition in the Skyrme case. This transition density decreases as the magnetic field grows \cite{prc-ang}. In fact, the results beyond this density are not physically meaningful but we show them for the sake of completeness in the density range considered in this work.
In Fig.~\ref{Fig5} contributions to the isothermal compressibility for both spin components in neutron matter are shown as a function of the logarithm of the magnetic field strength at saturation density for the Skyrme SLy7 (solid line) and Gogny D1P (dashed line) interactions. For each model upper (lower) curves refer to spin down (up) polarized particles. As the magnetic field strength increases there is a splitting of the spin up and down component behaviour with respect to the unpolarized case $(\Delta=0)$. Global compressibility is obtained by adding the contributions of both spin fractions and changes very mildly with the magnetic field. The relative variation of both components at the maximum field strength is of $20\%$ $(28\%)$ for the Skyrme SLy7 (Gogny D1P) interaction at this density. 

\begin{figure}[hbtp]
\begin{center}
\includegraphics [angle=-90,scale=.75] {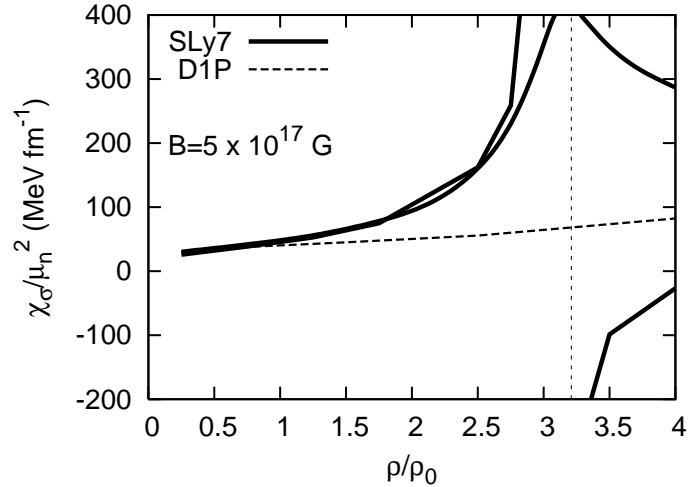}
\caption{Magnetic susceptibility as a function of density for the Skyrme SLy7 (solid line) and Gogny D1P (dashed line) models. See text for details.} 
\label{Fig6}
\end{center}
\end{figure}

We now consider the effect of a strong magnetic field on the static magnetic susceptibility. In Fig.~\ref{Fig6} we plot the magnetic susceptibility for neutron matter as a function of density in units of the nuclear magnetic moment squared, $\mu^2_n$, for a magnetic field $B=5 \times 10^{17}$ G. The results obtained with Skyrme SLy7 (Gogny D1P) interaction are shown with solid (dashed) line. For Skyrme results we can see that the onset of a ferromagnetic transition (that for this specific parametrization takes place at a density around $\rho=3.3\rho_0$ and is signaled with a vertical dotted line on the plot) drives a divergence of the susceptibility and the subsequent second order phase transition. In this model the upper curve corresponds to the $\chi^-/\mu^2_n$ value and the lower curve to the $\chi^+/\mu^2_n$. The onset of the ferromagnetic transition is due to the vanishing values approached by the quantities in the denominator of Eq.(\ref{chi}). For the Skyrme SLy7 interaction the meaningful regime, though, is limited for values before the transition takes place. For the Gogny D1P interaction (dashed line) the splitting between the contributions from the two spin orientations is less than $1\%$, therefore, indistinguishable on the plot. There is no ferromagnetic transition for this parametrization and the susceptibility keeps finite values for the whole density range.

\begin{figure}[hbtp]
\begin{center}
\includegraphics [angle=-90,scale=.75] {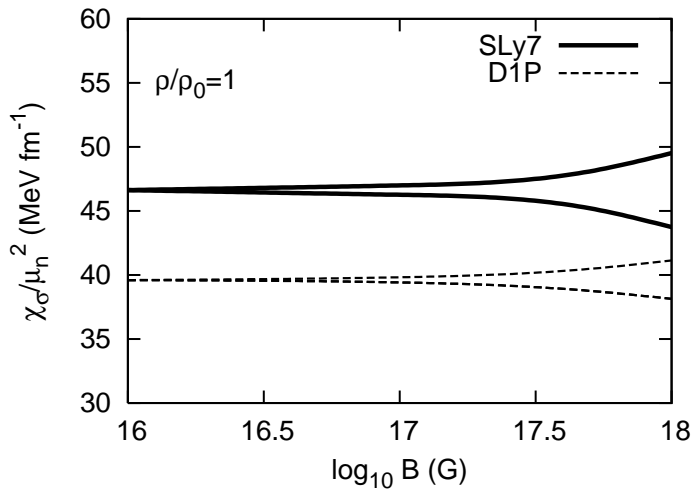}
\caption{Magnetic susceptibility for a pure neutron system as a function of the logarithm of the magnetic field strength for the Skyrme SLy7 (solid line) and Gogny D1P (dashed line) at saturation density. For each model upper (lower) curves refer to down (up) spin polarized components. See text for details.} 
\label{Fig7}
\end{center}
\end{figure}

In Fig.~\ref{Fig7} we show the magnetic susceptibility in units of the nuclear magnetic moment squared, $\mu^2_n$, for  neutron matter as a function of the logarithm of the magnetic field strength at saturation density. We plot results obtained with the Skyrme SLy7 (Gogny D1P) with solid (dashed) line. Upper (lower) curves for each model refer to spin down  (up) polarized population fractions. We see that contributions from the down (up) polarized components give larger (smaller) contribution to the total susceptibility that can be obtained as the sum of both. Notice that the global susceptibility remains almost unchanged as magnetic field grows. At this density the Skyrme model predicts a higher  tendency of the system  to achieve a net magnetization than in the Gogny case as $B$ increases. No transition is near at this density for the Skyrme model and a smooth behaviour is observed in the full range of variation of the magnetic field.

\section{Summary and conclusions}
\label{summary}
In this work we have investigated in the context of the Landau Theory of normal Fermi Liquids, the variation of monopolar $(l=0)$ and dipolar $(l=1)$ Landau parameters which describe the particle-hole interaction matrix elements for pure neutron matter in the presence of a strong magnetic field at zero temperature. We have used effective nuclear interactions such as the zero-range Skryme SLy7 and finite-range Gogny D1P and obtained their analytical expressions valid in the case of a polarized Fermi sea. We have computed the ratios of variation of these coefficients with respect to the $B=0$ case for the dominant spin component in the system. For the Skyrme and Gogny interactions the variation is very mild keeping below $|R|\approx 20\%$ for the maximum magnetic field strength studied in this work $B \approx 10^{18}$ G. This is a direct consequence of the small polarization induced by such strong fields in the system in the meaningful range of densities considered in this work.

We have also analyzed the effect of the presence of a strong magnetic field in some other static properties in the neutron plasma as deduced from the Landau Theory of Fermi Liquids. Effective neutron masses at the Fermi surface for each spin polarized component are calculated for both interactions from the dipolar $F_1$ coefficients. The magnetization in the plasma causes a splitting in the values of the up and down spin polarized components. Spin down (up) polarized particles show an increase (decrease) in their effective masses with respect to the non polarized case as the magnetic field strength grows. This will affect the level densities in polarized neutron matter. In addition, Skyrme interactions predict a larger variation in the splitting than the Gogny forces with increasing magnetic field, $B$. 
Other magnitudes such as the isothermal compressibility, $K$, relate to the dipolar Landau coefficients. In the presence of a strong magnetic field the compressibility is a growing function of density, stiffer in the Skyrme case than in the Gogny case. When a ferromagnetic transition is near, as it happens for the Skyrme interaction used in this work (as a representative case of almost all Skyrme parametrizations \cite{jerome}), there is a dramatic decrease of the compressibility around  the density of the onset of the ferromagnetic instability. The magnetic susceptibility, $\chi$, shows a divergent behaviour at densities close to that of the onset of the phase transion for the Skyrme case. The contribution to the total susceptibility from the polarized populations in the system is related to the  Landau coefficients in the spin channel that approach values driving the system to an instability. For the Gogny case there is no divergence since magnetization of the system keeps very mild. For densities in the intermediate range, not close to the transition, spin down (up) contributions give a higher (lower) susceptibility as it is more (less) easy to polarize the  system as the external field strengthens. 
Although a lot of work has been devoted to the study of polarized nuclear systems, additional effort should be made for exploring properties of nuclear asymmetric matter at finite temperature. The presence of magnetic fields should be carefully studied in the future with direct simulation techniques for the low density case. This will allow to  explore the possibilities that exotic matter shapes in the crust of neutron stars may give as an additional source of opacity to the neutrino cooling in the early stages of Supernovae cooling. 

\vspace{2ex}

\noindent{\bf Acknowledgments}\\

This work has been partially funded by the Spanish Ministry of Science and Innovation  (Micinn) under projects DGI-FIS2006-05319, FIS2007-60133, FIS2008-01661 and CPAN CSD2007-00042 Programa Consolider-Ingenio 2010.


\end{document}